\pdfoutput=1

\documentclass[runningheads]{llncs}
\usepackage[T1]{fontenc}
\usepackage{graphicx}
\usepackage{amsmath,amssymb,amsfonts}
\begin{document}
\title{Stretched sinograms for limited-angle tomographic reconstruction with neural networks}
\titlerunning{Stretched sinograms for neural tomographic reconstruction}
\author{Kyle Luther\inst{1} \and
H. Sebastian Seung\inst{1,2}}
\institute{Princeton Neuroscience Institute \and
Princeton Computer Science Department \\\email{kluther@princeton.edu}
}
\maketitle
\begin{abstract}



We present a direct method for limited angle tomographic reconstruction using convolutional networks. The key to our method is to first stretch every tilt view in the direction perpendicular to the tilt axis by the secant of the tilt angle. These stretched views are then fed into a 2-D U-Net which directly outputs the 3-D reconstruction. We train our networks by minimizing the mean squared error between the network's generated reconstruction and a ground truth 3-D volume. To demonstrate and evaluate our method, we synthesize tilt views from a 3-D image of fly brain tissue acquired with Focused Ion Beam Scanning Electron Microscopy. We compare our method to using a U-Net to directly reconstruct the unstretched tilt views and show that this simple stretching procedure leads to significantly better reconstructions. We also compare to using a network to clean up reconstructions generated by backprojection and filtered backprojection, and find that this simple stretching procedure also gives lower mean squared error on previously unseen images.
\end{abstract}

\section{Introduction}

Electron tomography is an imaging technique that uses transmission electron micrsope images acquired from multiple viewpoints to reconstruct the 3-D structure of an object \cite{frank1992electron}. Linear reconstruction techniques like filtered backprojection (FBP) or iterative reconstruction methods (SART,SIRT) are widely used but display strong artifacts in the presence of limited tilt angles, highly noisy inputs, and complex nonlinear misalignments, all of which are commonplace in electron tomography workflows \cite{natterer2001mathematics,zheng2022aretomo,mastronarde2017automated}.

\textbf{Problem setting} Our goal is to use deep learning to perform reconstruction of electron tomography tilt series. Ideally the implicit prior contained in a neural network will enable higher quality reconstructions using fewer tilts than would be required by classical reconstruction methods like filtered backprojection.

\textbf{Related methods} 
The related fields of X-Ray and positron emission tomography have seen a proliferation of deep learning methods to perform tomographic reconstruction. \cite{wang2020deep} categorize and review several distinct ways that neural networks have been used to aid in tomographic reconstruction of X-Ray CT data and we review the most relevant methods below.

\textit{Domain transform} methods use a neural network $f$ to map sinogram data $\mathbf{y}$ to a reconstruction $\hat{\mathbf{x}} = f(\mathbf{y})$. Perhaps the simplest approach from this category is the work of \cite{haggstrom2019deeppet} which uses a convolutional network to directly map sinograms to reconstructions. A related approach was used by \cite{whiteley2020directpet} which used hybrid fully connected/convolutional networks. \cite{zhu2018image} used fully connected layers and added an additional manifold encoding-decoding strategy. 

Notably these approaches have not shown the same level of interest as competing methods which we review below and \cite{ye2018deep} argue that the generalization performance of domain transform methods has been lackluster in addition to them typically having rather large computational and memory requirements.

\textit{Image domain} methods apply neural networks to the output of a classic reconstruction method, typically either backprojection or filtered backprojection. These methods can be thought of as neural post-processing. \cite{kang2017deep,chen2017low,shan2019competitive} demonstrate that deep networks can improve the quality of reconstructions that have been generated from low-dose (noisy) views. \cite{jin2017deep} show that artifacts caused by reconstruction from sparse viewpoints can be removed, and \cite{bubba2019learning} show that artifacts caused by a limited-angle set of views can be removed using deep networks to post-process reconstructions. \cite{zhang2018convolutional} show that post-processing can be even used to remove metal artifacts.
A hybrid method was used in \cite{ye2018deep,li2021sparse} where the first step was to map a sinogram of size nview $\times$ width to a sequence of individual backprojections of size nview $\times$ height $\times$ width. 

\textit{Sensor domain} methods instead apply neural networks to the raw sinograms and are typically only used as a pre-processing step so that classical reconstruction methods are applied to the network outputs. Networks have been shown to remove artifacts \cite{claus2017metal,ghani2019fast}. Networks have also been used to inpaint missing views from limited-angle sinograms \cite{tovey2019directional}. 

\textit{Dual domain} methods combine both \textit{sensor} and \textit{image} domain approaches and apply networks both before and after backprojection \cite{lin2019dudonet,wang2021improving,wang2022dudotrans}. 

\textit{Dictionary-based reconstruction of affine-aligned tilt views} On a seemingly unrelated front, dictionary based reconstruction with sparsity priors was applied to electron tomographic reconstruction \cite{hu2013electron,veeraraghavan2010increasing}. Critically, these works first performed an \textit{affine} alignment to tilt views before reconstruction. Using an affine alignment is non-standard as it stretches every view instead of just translating them. However, both works showed impressive reconstructions when combining the affine-stretched views and a translationally invariant dictionary. 

\textbf{What is missing?} With the exception of \cite{haggstrom2019deeppet}, prior neural methods either rely on fully connected architectures \cite{whiteley2020directpet,zhu2018image} or backprojection to actually perform the reconstruction. Our goal is generate quality reconstructions with the widely used U-Net architecture \cite{ronneberger2015unet} and avoid the use of backprojection.

\textbf{Our contributions} 
We propose a simple backprojection-free pre-processing scheme that significantly improves the ability of a convolutional network to perform tomographic reconstructions. We simply stretch each view in the sinogram by $\sec\theta$ along the direction perpendicular to the tilt axis. Our method is applicable to parallel beam, limited angle geometries that are typical in electron tomography.

\section{Method: neural reconstruction of stretched sinograms}
\begin{figure}
    \centering
    \includegraphics[width=\linewidth]{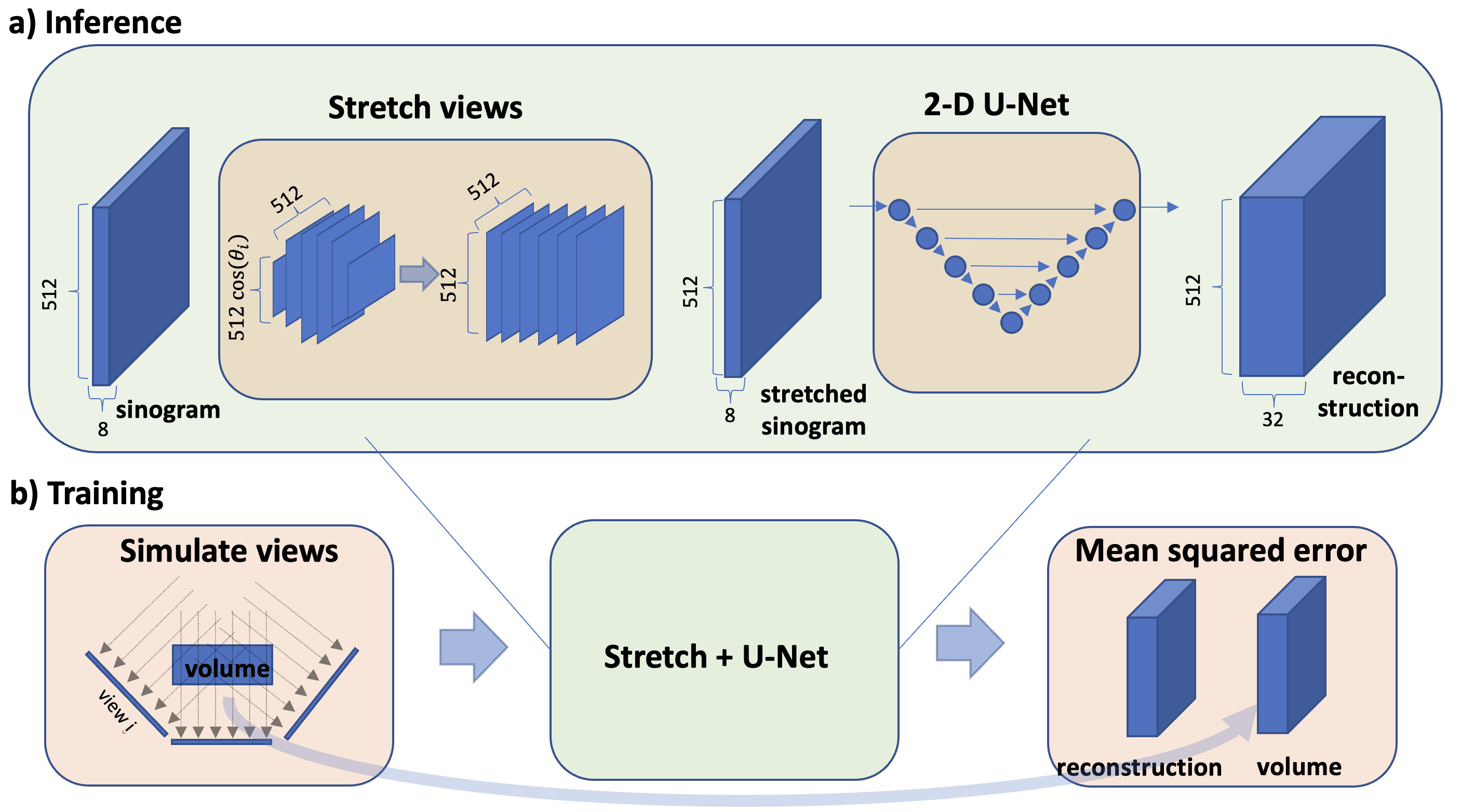}
    \caption{Method overview \textbf{Inference}: we first stretch every tilt view by $\sec\theta$ in the direction perpendicular to the tilt axis, then perform reconstruction with a U-Net \textbf{Training}: We train our networks using simulated tilt views and mean squared error (MSE) between network outputs and ground truth volumes }
    \label{fig:overview}
\end{figure}
Our training and reconstruction procedures are shown in Fig. \ref{fig:overview}. 
\subsection{Stretched sinograms} 
\textbf{Notation and assumptions} We assume we are working with a parallel beam geometry and that we have $n_{\text{view}}$ tilt views of size $n_h \times n_w$ over a limited range of angles $(-\theta, +\theta)$ where $\theta < 90^\circ$. We denote the 3-D tensor of 2-D tilt views by $\mathbf{y} \in \mathbb{R}^{n_{\text{view}} \times n_h \times n_w}$. We assume these views have been log-normalized so that they are related to the object density $\mathbf{x} \in \mathbb{R}^{n_d \times n_h \times n_w}$ via the Radon transform:
\begin{equation}
    \mathbf{y} = P \mathbf{x}
\end{equation}

\textbf{Generating stretched views} To generate the stretched sinogram, we simply stretch every tilt view by $\sec\theta$ in the direction perpendicular to the tilt axis. This stretching is performed so that the image size is preserved, meaning the stretched sinograms are also $n_h \times n_w$ in size. Bilinear interpolation is used to perform the stretching and the stretching is done treating the center of each image as the origin. 

If the tilt axis is parallel to the $\hat{y}$ axis, this means we stretch along the $\hat{x}$-axis of each 2D tilt view. We now write the formula for tilt stretching in this case. Let $i,j$ index the rows,columns of a tilt image. We use natural coordinates so $(-1,-1)$ refers to the upper left and $(+1,+1)$ refers to the bottom right corner of the image. Bilinear interpolation is used to evaluate at fractional pixel locations.
\begin{equation}
    y^{\textrm{stretched}}(\theta, i, j) = y\left(\theta,i, j \sec\theta\right)
\label{eqn:stretch_indexform}
\end{equation}
This simple stretching procedure can be extended to stacks with arbitrary tilt axes (e.g. dual tilt setups \cite{mastronarde1997dual}). The formula in Eq. \ref{eqn:stretch_indexform} would need change to stretch along a linear combination of $i,j$. We can write the stretching procedure in matrix notation, treating bilinear interpolation as a sparse linear operator $S$:
\begin{equation}
    \mathbf{y}^{\textrm{stretched}} = S \mathbf{y}
\label{eqn:stretch_matrixform}
\end{equation}
This operation maps an $n_{\text{view}} \times n_{h} \times n_{w}$ tensor of raw views to an $n_{\text{view}} \times n_h \times n_{w}$ tensor of stretched views.

\begin{figure}
    \centering
    \includegraphics[width=\linewidth]{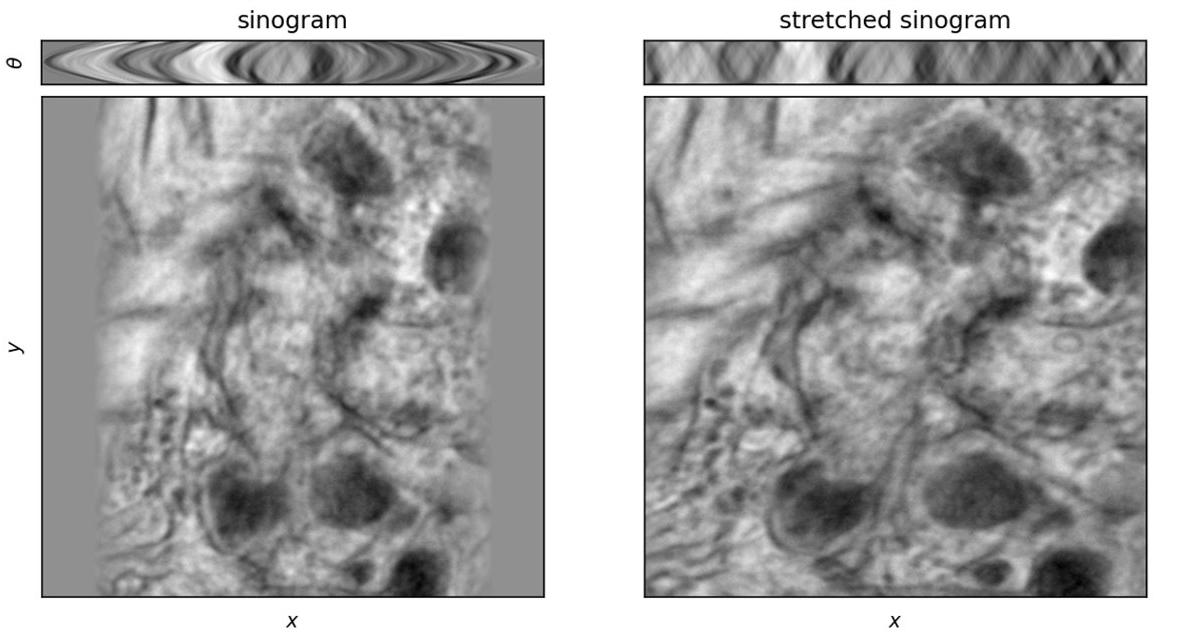}
    \caption{Sinogram vs stretched sinogram representations. The sinogram is generated with a vertical tilt axis and with tilt angles spaced at $2^\circ$ increments between $-45^\circ$ and $+45^\circ$. The stretched sinogram is by stretching each $xy$ view of the sinogram by $\sec\theta$ (Eq. \ref{eqn:stretch_indexform}).
    In the top row we show an $x \theta$-slice of the raw/stretched sinograms. In the bottom row we show $xy$ slices corresponding to $\theta=-45^\circ$.}
    \label{fig:stretched_sinogram}
\end{figure}
\textbf{Visualizing the stretched views} In Fig. \ref{fig:stretched_sinogram} we compare a simulated sinogram and its corresponding stretched sinogram. We simulate the sinogram by applying the Radon transform to a $45 \times 512 \times 512$ volume of FIB-SEM data of fly brain tissue \cite{xu2020connectome,scheffer2020connectome} which is further described in Section 3. Specifically we compute 45 projections uniformly spaced over the angles $(-45^\circ,+45^\circ)$ with a tilt axis in the $\hat{y}$ direction. We show the $x\theta$ view for $y=1$ and the $xy$ view for $\theta=-45^\circ$ for both the sinogram and stretched sinogram.

\subsection{3-D Reconstruction with a 2-D U-Net}
For simplicity we stick closely to the original 2-D U-Net architecture proposed in \cite{ronneberger2015unet}. More details are provided in the Appendix. We treat the stretched sinogram with $K$ tilt views as a 2-D image with $K$ input channels. We generate a 3-D reconstruction from the 2-D U-Net by treating the output channel dimension as the depth dimension of the output volume. To improve training speed we use Instance Normalization layers before every ReLU layer \cite{ulyanov2016instance}. We found that the more popular Batch Normalization operation gave significantly higher test time error, likely due to the fact that we train our U-Nets with a batch size of 1.

\subsection{Supervised network training}
We assume we have paired examples of tilts $\mathbf{y}$ and volumes $\mathbf{x}$. We train our networks using stochastic gradient descent applied to the mean squared error (MSE) objective between these volumes and network reconstructions. Specifically, we generate the network reconstruction $\hat{\mathbf{x}}=f_{\omega} (S\mathbf{y})$ using a U-Net applied to the stretched tilt views. We then compute and backpropagate through the MSE:
\begin{equation}
    l(\omega) = \frac{1}{n_{oixel}} \left \Vert \mathbf{x} - f_{\omega} (S\mathbf{y}) \right \Vert^2 
\end{equation}
where $n_{pixel} = n_d \times n_h \times n_w$ is the number of pixels in the output reconstructions.

\section{Experiments with simulated data}
\begin{figure}
    \centering
    \includegraphics[width=\linewidth]{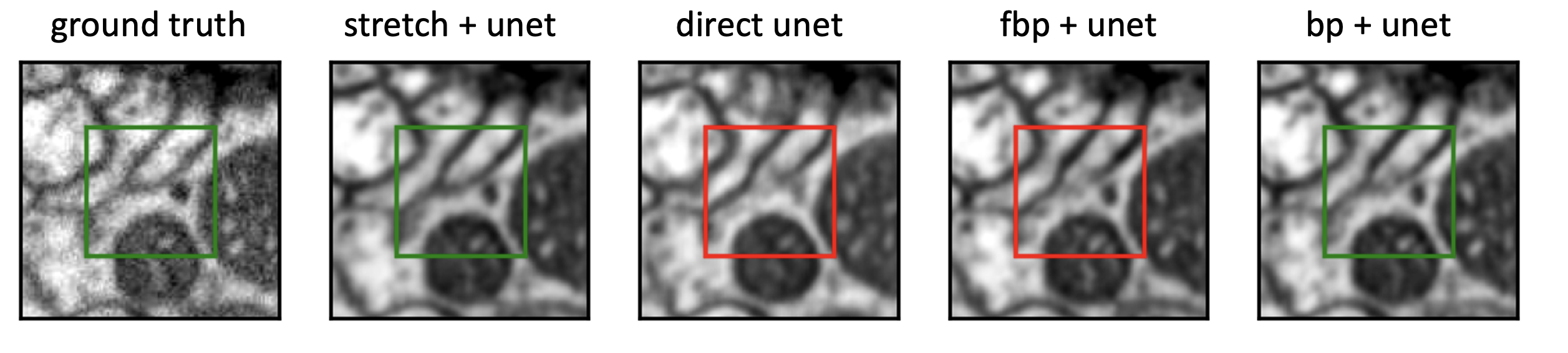}
    \caption{$128 \times 128$ sized crop from an $xy$ view of a test set reconstruction. Directly using a U-Net to perform reconstruction or applying a U-Net to clean up FBP causes a neuron boundary to disappear. }
    \label{fig:qualitative_results}
\end{figure}

\subsection{Data and simulated tilt views}
We download three $1k \times 1k \times 1k$ voxel volumes to use as the train, validation, and test sets from the 3D isotropic dataset imaged at $8 \times 8 \times 8$ nm$^3$ resolution by \cite{xu2020connectome}. This dataset contains images of a fruit-fly brain imaged with a focused ion beam milling and scanning electron microscope imaging (FIB-SEM) technique. Slices from this dataset are shown in the Appendix.

We sample patches of size $32 \times 512 \times 512$ voxels at arbitrary positions from the $1k \times 1k \times 1k$ voxel training volume. These patches are normalized to be zero-mean, unit-variance. From these patches, simulated tilt series consisting of 8 tilt views ranging uniformly over $(-60^\circ,+60^\circ)$ are generated using the ASTRA toolbox \cite{van2015astra,van2016fast}. The input shape to our networks is therefore $8 \times 512 \times 512$.

For data augmentation we add Gaussian noise with standard deviation of $0.3$ times the standard deviation of the projections and randomly shift a chosen number $n$ of the 8 tilt views in both $x$ and $y$ directions by an integer number of pixels between $[-3,+3]$ to simulate misalignments that can occur in real tomographic tilt series. We compare results for when we misalign $n=0,2,4,8$ of the sections. Each tilt view is then normalized to be unit mean and zero variance after augmentation.

\subsection{Comparisons}
We compare our method representation to 5 other methods: (1) using the raw sinograms as input to a U-Net, (2) using backprojection as input to a U-Net, (3) using filtered backprojection as input to a U-Net (the approach of \cite{jin2017deep}), (4) using backprojection without U-Net postprocessing, (5) using filtered backprojection without U-Net postprocessing. We use the ASTRA Toolbox to implement both the backprojection and filtered backprojection operations. For filtered backprojection, we use the default \textit{ram-lak} filter which implements a ramp-function in frequency space. In the case of backprojection and filtered-backprojection, the depth dimension is treated as the channel index to the 2D U-Nets and the channel dimensionality is $32$ instead of $8$. 

\subsection{Training details}
The forward and backward projections were implemented on a CPU using the ASTRA toolbox and these were distributed across 64 cpus to speed up inference time. We use a single ASUS RTX3090 Turbo 24GB GPU for inference and backpropagation. Networks were trained using PyTorch \cite{paszke2019pytorch}. We use the Adam optimizer to train our U-Nets \cite{kingma2014adam}. We trained all networks for up to 72 hours with a step-size of $0.001$. Learning curves for all configurations are shown in the Appendix. We used early stopping computed on the validation set so that we perform test-set evaluation using the network with lowest validation set error. 

\subsection{Results}
In Fig. \ref{fig:qualitative_results} we show a qualitative comparison between various methods for one $xy$ slice of the reconstruction on the test set (in the setting where 4/8 of the tilt views were misaligned). We evaluate our networks on a $ 992 \times 1000 \times 1000 $ test set volume extracted from a neighboring location of FIB-SEM images of fly brain tissue. This volume is broken into 124 (almost) non-overlapping patches\footnote{There is a small sliver of width 24 pixels where the patches overlap due to the fact that we are extracting $512 \times 512$ patches from $1000 \times 1000$ sections} of size $32 \times 512 \times 512$ and tilt views are simulated and subsequently reconstructed.

\textbf{Vary level of noise}
\begin{table}[]
\centering
\begin{tabular}{| c | c | c | c | c | c | c |}
    \hline
    noise & direct U-Net & stretch+U-Net & fbp+U-Net & bp+U-Net & fbp & bp\\
    \hline
    $0.0$ & $0.145 \pm 0.001$ & $\mathbf{0.092 \pm 0.001}$ & $0.120 \pm 0.001$ & $0.097 \pm 0.001$ & $0.389 \pm 0.003 $ &  $0.451 \pm 0.005 $ \\
    $0.1$ & $0.146 \pm 0.002$ & $\mathbf{0.093 \pm 0.001}$ & $0.119 \pm 0.001$ & $0.098 \pm 0.001$ & $0.637 \pm 0.003 $ &  $0.452 \pm 0.005 $ \\
    $0.2$ & $0.150 \pm 0.002$ & $\mathbf{0.098 \pm 0.001}$ & $0.122 \pm 0.001$ & $0.103 \pm 0.001$ & $0.996 \pm 0.004 $ &  $0.454 \pm 0.005 $ \\
    $0.3$ & $0.157 \pm 0.002$ & $\mathbf{0.104 \pm 0.001}$ & $0.129 \pm 0.001$ & $0.109 \pm 0.001$ & $1.244 \pm 0.004 $ &  $0.458 \pm 0.005 $ \\
    $0.4$ & $0.166 \pm 0.002$ & $\mathbf{0.114 \pm 0.001}$ & $0.140 \pm 0.001$ & $0.119 \pm 0.001$ & $1.403 \pm 0.004 $ &  $0.462 \pm 0.005 $ \\    \hline
\end{tabular}
\caption{Test set MSE as we vary the level of noise added to tilts at inference time. In each setting we reconstruct $32 \times 512 \times 512$ volumes from $8$ tilt views of size $512 \times 512$ with tilt angles uniformly spaced between $[-60^\circ,+60^\circ]$.}
\label{tab:results_noise}
\end{table}
In this setup we use networks trained on non-misaligned data and do not misalign tilts at inference time either. Here we vary the level of noise at inference time. We observe that as we add noise, the MSE of plain filtered backprojection dramatically increases. All neural methods improve over both classical methods. We observe that not stretching the inputs before inference with a U-Net gives nearly $50 \%$ higher mean squared error than our method. We observe that stretching provides the lowest MSE with \textit{bp+U-Net} closesly following behind. The results are shown in Tab. \ref{tab:results_noise}.

\textbf{Vary level of misalignments}
\begin{table}[]
\centering
\begin{tabular}{| c | c | c | c | c | c | c |}
    \hline
    n shifts & direct U-Net & stretch+U-Net & fbp+U-Net & bp+U-Net & fbp & bp\\
    \hline
    0/8 & $0.157\pm 0.002$ & $\mathbf{0.104\pm 0.001}$ & $0.129\pm 0.001$ & $0.109\pm 0.001$ & $1.244\pm 0.004$ & $0.458 \pm 0.005$ \\
    2/8 & $0.217 \pm 0.002$ & $\mathbf{0.118 \pm 0.001}$ & $ 0.157 \pm 0.001$ & $0.130 \pm 0.001$ & $1.274 \pm 0.004 $ & $0.479 \pm 0.001$ \\
    4/8 & $ 0.226 \pm 0.003$ & $\mathbf{ 0.121 \pm 0.001}$ & $ 0.166 \pm 0.002$ & $0.135 \pm 0.001$ & $1.304 \pm 0.004$ & $0.481 \pm 0.005$ \\
    8/8 & $0.235 \pm 0.002$ & $\mathbf{0.126 \pm 0.001}$ & $0.180 \pm 0.001$ & $0.164 \pm 0.002$ & $1.367 \pm 0.004$ & $0.508 \pm 0.006$ \\
    \hline
\end{tabular}
\caption{Test set MSE as we vary the number of misaligned tilt views (both at train and test time). In each setting we reconstruct $32 \times 512 \times 512$ volumes from $8$ tilt views of size $512 \times 512$ with tilt angles uniformly spaced between $[-60^\circ,+60^\circ]$.}
\label{tab:results_misalign}
\end{table}

In this setup we train and test networks with various number of misaligned tilts. All neural methods improve over both classical methods. We observe that stretching the tilts before inference gives much lower MSE than not stretching. This time, we observe a larger gap between \textit{stretch+U-Net} compared to \textit{bp+U-Net} or \textit{fbp+U-Net} as we increase the number of misaligned sections. The results are shown in Tab. \ref{tab:results_misalign}.

\section{Discussion}

Why should using the stretched sinogram as network input lead to lower reconstruction error than backprojection or filtered backprojection? Both BP and FBP are reconstruction algorithms so intuitively it might seem like cleaning up a reconstruction would be easier to generating one from scratch like networks do in our method. Our experiments suggest however that this intuition is wrong. We speculate that it is because backprojection \textit{attenuates} high frequency modes while filtered backprojection \textit{amplifies} in the input. Stretching on the other hand, may suppress less information in the input which may be helpful for data-driven learning.

\textbf{Limitations} In this proof of concept, our tilt views were synthesized from a 3-D image, which could serve as ground truth for the desired reconstruction. In the real world, where would the ground truth for supervised training come from? One possible scenario is that conventional tomography would be used to generate the ground truth for convolutional net tomography. One would first acquire a high quality, densely sampled, low noise, well-aligned set of tilt views. These views given to classic reconstruction methods like FBP or SIRT to generate 3D reconstructions. Neural nets would then be trained to perform tomography from subsets of the tilt views (possibly with additional noise and misalignment augmentations). Once trained, the nets could be applied in large scale tomographic imaging pipelines, where there is a strong motivation to reduce the number of tilt views and/or acquisition time of each image.

We have relied on mean squared error as the primary metric for evaluating and comparing methods. In cell biology, the goal is to learn something new about the structure of the object being reconstructed. In other fields like connectomics, a central goal is identify boundaries between cells \cite{dorkenwald2022flywire}. Neither of these goals is directly linked to MSE. We have attempted to validate our reconstruction via qualitative inspections and have indeed seen cases where membranes (in particular membranes that lie in the imaging plane) are washed out. This suggests that downstream segmentation of neurites may indeed benefit via stretched sinogram representations. However this point must be remembered when interpreting the results of this study.

\textbf{Extensions} As discussed in the limitations section a central goal is to use tomography as part of a broader pipeline. Many pipelines in the literature already use deep neural networks to generate segmentations of electron microscope images. It seems reasonable to do away with explicit reconstruction altogether and instead directly map tilt views to segmentation. This has the bonus of avoiding any questions regarding MSE, since the network would be directly optimizing the quantity of interest: namely the segmentation performance.

\section*{Appendix}
\subsection*{Dataset visualization}
\begin{figure}[h]
    \centering
    \includegraphics[width=\linewidth]{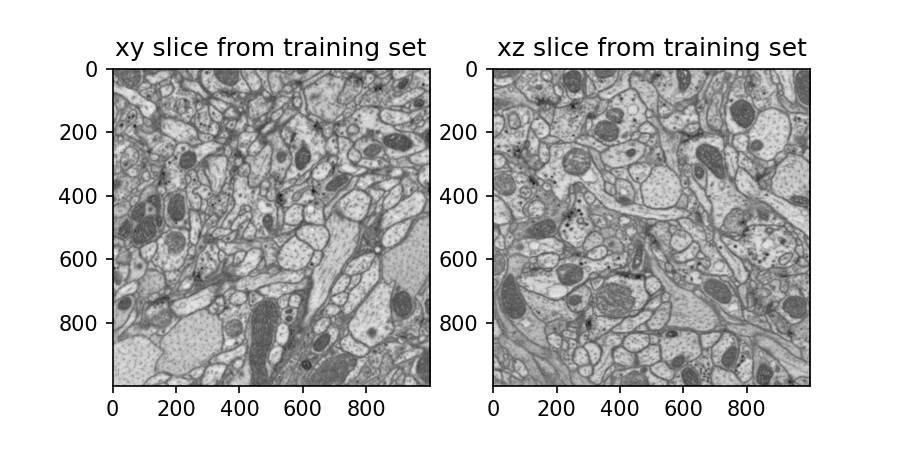}
    \caption{(left) 1 of the 1000 xy slices from the training set (right) 1 of the 1000 xz slices from the training set. 32 x 512 x 512 patches at arbitrary zyx locations are sampled from this volume during training.}
    \label{fig:data_overiew}
\end{figure}

In Fig. \ref{fig:data_overiew} we show an example xy and xz from the training dataset.

The datasource can be publicly viewed at \textit{https://hemibrain-dot-neuroglancer-demo.appspot.com}. The bounding box for the training dataset is defined by xyz = (22000, 19000, 20000) to xyz = (23000, 20000, 21000). The bounding box for the validation dataset is defined by xyz = (22000, 21000, 20000) to xyz = (23000, 22000, 21000). The bounding box for the training dataset is defined by xyz = (22000, 23000, 20000) to xyz = (23000, 24000, 21000). 

\subsection*{U-Net architecture}
The number of feature maps at each of the 5 levels in our U-Net is the same as in \cite{ronneberger2015unet}. However, instead of valid padding, we use same-size padding so the input and output are both 512 x 512. We also use Instance Normalization layers before every ReLU layer \cite{ulyanov2016instance}.





\subsection*{Learning curves}
We show training and validation learning curves for each of the 4 x 4=16 configurations of networks trained. In all settings, a learning rate of 0.001 with the Adam optimizer was applied. The curves are shown for 72 hours of single gpu training. Fewer steps were taken for backprojection and filtered backprojection as these operations were performed on CPU and took a significant amount of time.

The labels on the graphs (sinogram, stretched sinogram, filtered backprojection, backprojection) refer to the preprocessing method used before inputting to the neural network. To generate our evaluations in the main text (Table 1 and 2), early stopping was used. This means we used the network at the iteration with lowest validation loss (rather than the network at iteration 2M or 2.4M).

It is interesting to note that for all degrees of misalignment, the minimum validation loss is achieved after many more iterations for the backprojection and filtered backprojection inputs. Future studies will be needed to test if this can be changed by using a different learning rate for the BP/FBP inputs or if this difference reflects a fundamental behavior about training networks to clean up BP/FBP inputs.

\begin{figure}[h]
    \centering
    \includegraphics[width=\linewidth]{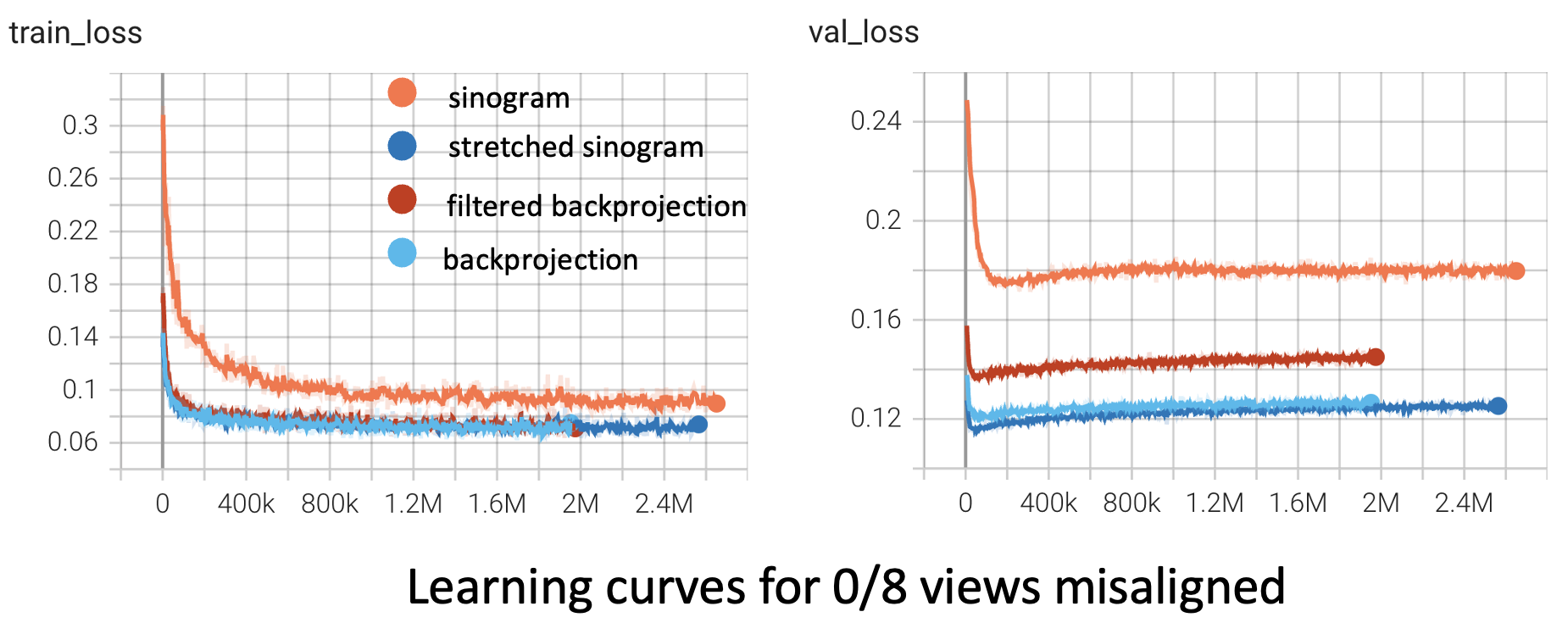}
    \caption{Learning curves for 0/8 misaligned views. Interestingly backprojection and filtered backprojection have very similar training losses, but notably different validation losses. In all settings, the loweset validation loss is achieved very early into training (50-100K training iterations)}
    \label{fig:lcurves_m0}
\end{figure}

\begin{figure}
    \centering
    \includegraphics[width=\linewidth]{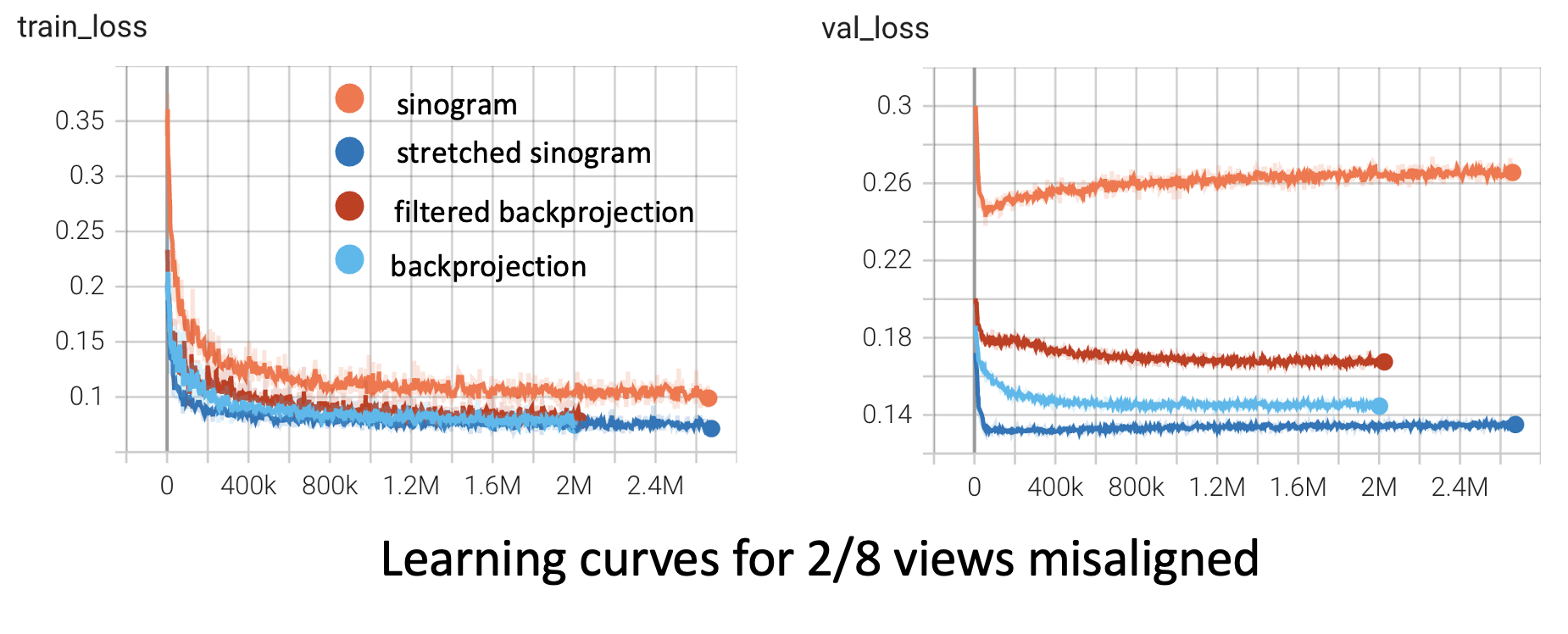}
    \caption{Learning curves for 2/8 misaligned views. Stretched sinogram, backprojection and filtered backprojection inputs to the UNet have very similar training losses, but notably different validation losses. This time backprojection and filtered backprojection validation curves take much longer than sinogram/stretched sinogram validation curves to flatten out.}
    \label{fig:lcurves_m2}
\end{figure}

\begin{figure}[h]
    \centering
    \includegraphics[width=\linewidth]{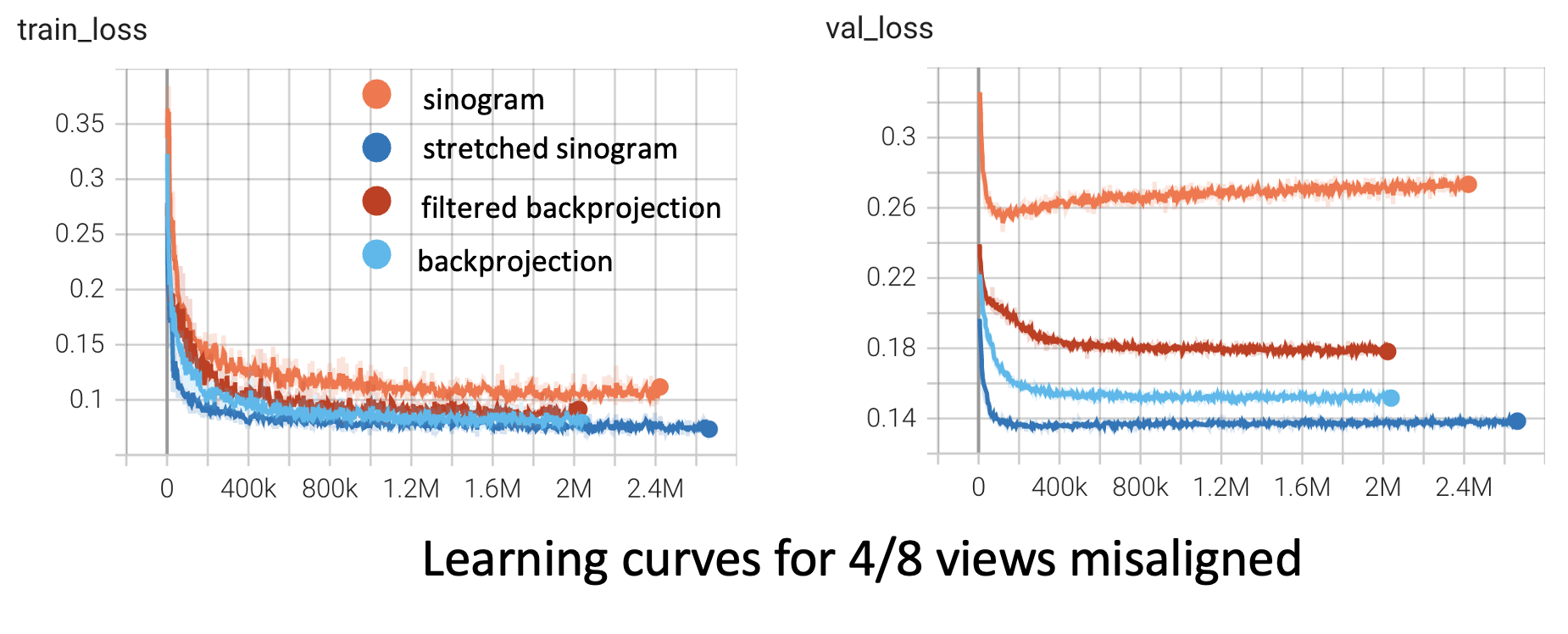}
    \caption{Learning curves for 4/8 misaligned views. These curves very closely resemble the curves in the setting where only 2/8 views are misaligned, expect they are shifted up (slightly higher loss for all configs).}
    \label{fig:lcurves_m4}
\end{figure}

\begin{figure}[h]
    \centering
    \includegraphics[width=\linewidth]{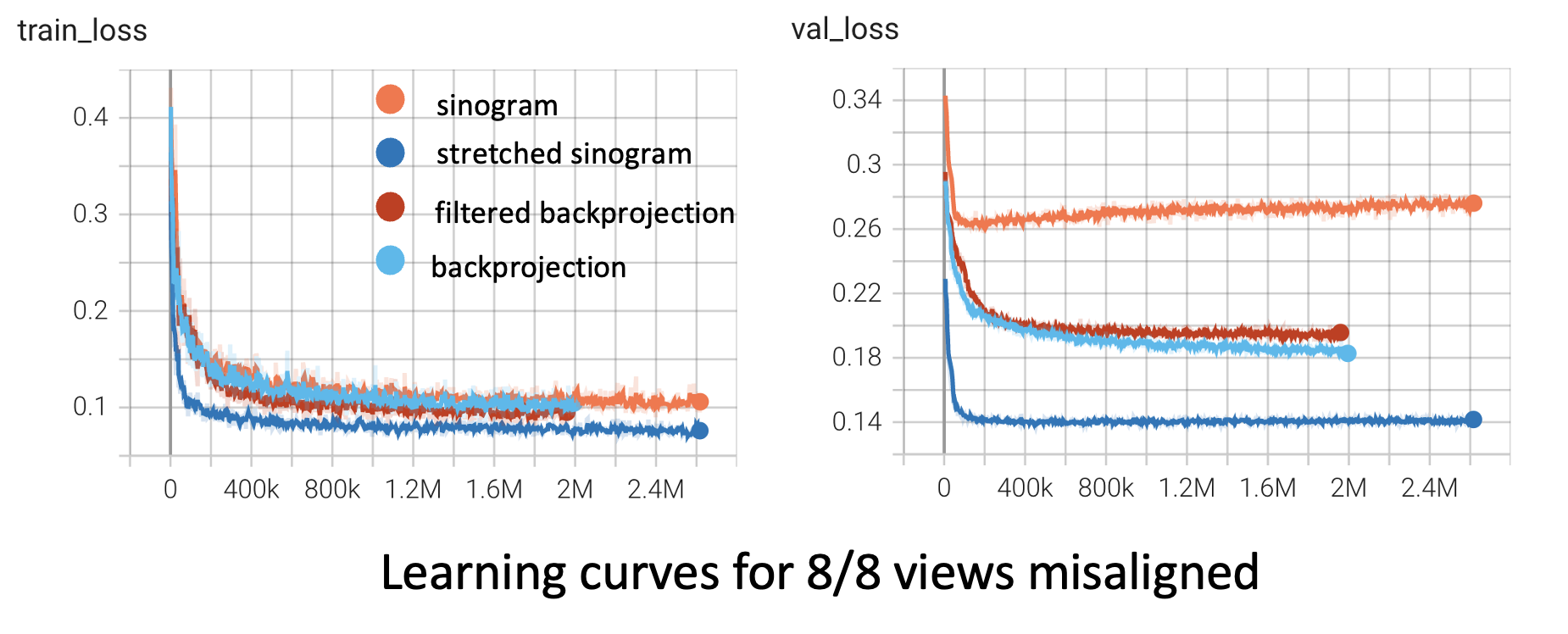}
    \caption{Learning curves for 8/8 misaligned views. Interestingly the sinogram and streetched sinogram validation curves still reach their minimum relatively early into training (100K for sinogram inputs and 400 K for stretched sinogram inputs). Now however the backprojection and filtered backprojection validation curves appear to continue to decrease much longer into their training trajectories.}
    \label{fig:lcurves_m8}
\end{figure}

\subsection*{Qualitative comparisons}
\begin{figure}
    \centering
    \includegraphics[width=\linewidth]{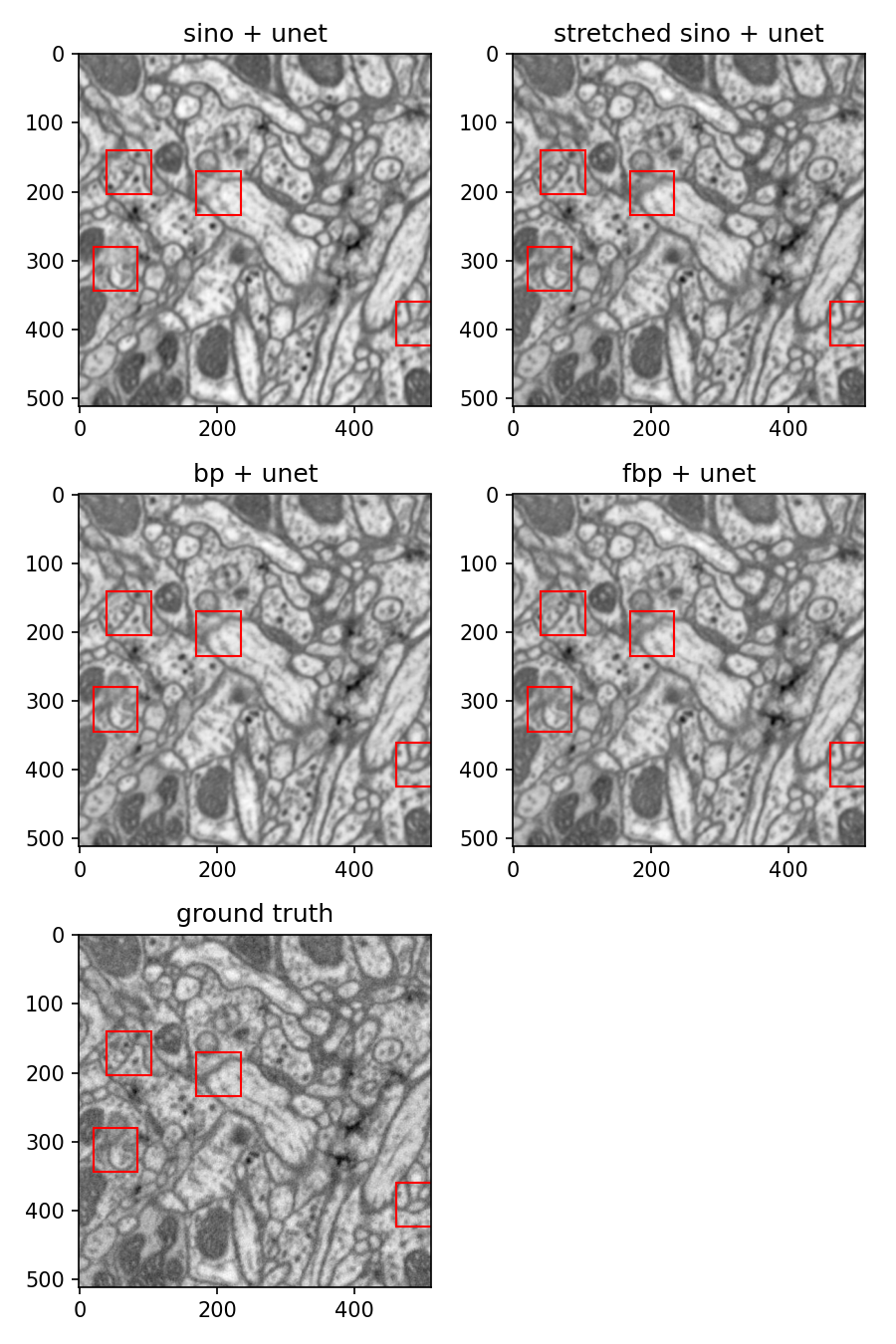}
    \caption{The central xy slice of test set reconstructions given in the setting with 0 misaligned tilt views. We also compare this with the ground truth. Red boxes highlight locations where boundaries are broken by at least one of the methods (usually the direct sinogram input).}
    \label{fig:qualitative0}
\end{figure}

In Fig. \ref{fig:qualitative0}, we show an xy slice of test set reconstructions given in the setting with 0 misaligned tilt views. We also compare this with the ground truth.

\begin{thebibliography}{10}
\providecommand{\url}[1]{\texttt{#1}}
\providecommand{\urlprefix}{URL }
\providecommand{\doi}[1]{https://doi.org/#1}

\bibitem{bubba2019learning}
Bubba, T.A., Kutyniok, G., Lassas, M., M{\"a}rz, M., Samek, W., Siltanen, S.,
  Srinivasan, V.: Learning the invisible: a hybrid deep learning-shearlet
  framework for limited angle computed tomography. Inverse Problems
  \textbf{35}(6),  064002 (2019)

\bibitem{chen2017low}
Chen, H., Zhang, Y., Zhang, W., Liao, P., Li, K., Zhou, J., Wang, G.: Low-dose
  ct via convolutional neural network. Biomedical optics express
  \textbf{8}(2),  679--694 (2017)

\bibitem{claus2017metal}
Claus, B.E., Jin, Y., Gjesteby, L.A., Wang, G., De~Man, B.: Metal-artifact
  reduction using deep-learning based sinogram completion: initial results. In:
  Proc. 14th Int. Meeting Fully Three-Dimensional Image Reconstruction Radiol.
  Nucl. Med. pp. 631--634 (2017)

\bibitem{dorkenwald2022flywire}
Dorkenwald, S., McKellar, C.E., Macrina, T., Kemnitz, N., Lee, K., Lu, R., Wu,
  J., Popovych, S., Mitchell, E., Nehoran, B., et~al.: Flywire: online
  community for whole-brain connectomics. Nature methods  \textbf{19}(1),
  119--128 (2022)

\bibitem{frank1992electron}
Frank, J., et~al.: Electron tomography. Springer (1992)

\bibitem{ghani2019fast}
Ghani, M.U., Karl, W.C.: Fast enhanced ct metal artifact reduction using data
  domain deep learning. IEEE Transactions on Computational Imaging  \textbf{6},
   181--193 (2019)

\bibitem{haggstrom2019deeppet}
H{\"a}ggstr{\"o}m, I., Schmidtlein, C.R., Campanella, G., Fuchs, T.J.: Deeppet:
  A deep encoder--decoder network for directly solving the pet image
  reconstruction inverse problem. Medical image analysis  \textbf{54},
  253--262 (2019)

\bibitem{hu2013electron}
Hu, T., Nunez-Iglesias, J., Vitaladevuni, S., Scheffer, L., Xu, S.,
  Bolorizadeh, M., Hess, H., Fetter, R., Chklovskii, D.B.: Electron microscopy
  reconstruction of brain structure using sparse representations over learned
  dictionaries. IEEE Transactions on Medical Imaging  \textbf{32}(12),
  2179--2188 (2013)

\bibitem{jin2017deep}
Jin, K.H., McCann, M.T., Froustey, E., Unser, M.: Deep convolutional neural
  network for inverse problems in imaging. IEEE Transactions on Image
  Processing  \textbf{26}(9),  4509--4522 (2017)

\bibitem{kang2017deep}
Kang, E., Min, J., Ye, J.C.: A deep convolutional neural network using
  directional wavelets for low-dose x-ray ct reconstruction. Medical physics
  \textbf{44}(10),  e360--e375 (2017)

\bibitem{kingma2014adam}
Kingma, D.P., Ba, J.: Adam: A method for stochastic optimization. arXiv
  preprint arXiv:1412.6980  (2014)

\bibitem{li2021sparse}
Li, W., Buzzard, G.T., Bouman, C.A.: Sparse-view ct reconstruction using
  recurrent stacked back projection. In: 2021 55th Asilomar Conference on
  Signals, Systems, and Computers. pp. 862--866. IEEE (2021)

\bibitem{lin2019dudonet}
Lin, W.A., Liao, H., Peng, C., Sun, X., Zhang, J., Luo, J., Chellappa, R.,
  Zhou, S.K.: Dudonet: Dual domain network for ct metal artifact reduction. In:
  Proceedings of the IEEE/CVF Conference on Computer Vision and Pattern
  Recognition. pp. 10512--10521 (2019)

\bibitem{mastronarde1997dual}
Mastronarde, D.N.: Dual-axis tomography: an approach with alignment methods
  that preserve resolution. Journal of structural biology  \textbf{120}(3),
  343--352 (1997)

\bibitem{mastronarde2017automated}
Mastronarde, D.N., Held, S.R.: Automated tilt series alignment and tomographic
  reconstruction in imod. Journal of structural biology  \textbf{197}(2),
  102--113 (2017)

\bibitem{natterer2001mathematics}
Natterer, F.: The mathematics of computerized tomography. SIAM (2001)

\bibitem{paszke2019pytorch}
Paszke, A., Gross, S., Massa, F., Lerer, A., Bradbury, J., Chanan, G., Killeen,
  T., Lin, Z., Gimelshein, N., Antiga, L., et~al.: Pytorch: An imperative
  style, high-performance deep learning library. Advances in neural information
  processing systems  \textbf{32} (2019)

\bibitem{ronneberger2015unet}
Ronneberger, O., Fischer, P., Brox, T.: U-net: Convolutional networks for
  biomedical image segmentation. In: Medical Image Computing and
  Computer-Assisted Intervention--MICCAI 2015: 18th International Conference,
  Munich, Germany, October 5-9, 2015, Proceedings, Part III 18. pp. 234--241.
  Springer (2015)

\bibitem{scheffer2020connectome}
Scheffer, L.K., Xu, C.S., Januszewski, M., Lu, Z., Takemura, S.y., Hayworth,
  K.J., Huang, G.B., Shinomiya, K., Maitlin-Shepard, J., Berg, S., et~al.: A
  connectome and analysis of the adult drosophila central brain. Elife
  \textbf{9},  e57443 (2020)

\bibitem{shan2019competitive}
Shan, H., Padole, A., Homayounieh, F., Kruger, U., Khera, R.D., Nitiwarangkul,
  C., Kalra, M.K., Wang, G.: Competitive performance of a modularized deep
  neural network compared to commercial algorithms for low-dose ct image
  reconstruction. Nature Machine Intelligence  \textbf{1}(6),  269--276 (2019)

\bibitem{tovey2019directional}
Tovey, R., Benning, M., Brune, C., Lagerwerf, M.J., Collins, S.M., Leary, R.K.,
  Midgley, P.A., Sch{\"o}nlieb, C.B.: Directional sinogram inpainting for
  limited angle tomography. Inverse problems  \textbf{35}(2),  024004 (2019)

\bibitem{ulyanov2016instance}
Ulyanov, D., Vedaldi, A., Lempitsky, V.: Instance normalization: The missing
  ingredient for fast stylization. arXiv preprint arXiv:1607.08022  (2016)

\bibitem{van2016fast}
Van~Aarle, W., Palenstijn, W.J., Cant, J., Janssens, E., Bleichrodt, F.,
  Dabravolski, A., De~Beenhouwer, J., Batenburg, K.J., Sijbers, J.: Fast and
  flexible x-ray tomography using the astra toolbox. Optics express
  \textbf{24}(22),  25129--25147 (2016)

\bibitem{van2015astra}
Van~Aarle, W., Palenstijn, W.J., De~Beenhouwer, J., Altantzis, T., Bals, S.,
  Batenburg, K.J., Sijbers, J.: The astra toolbox: A platform for advanced
  algorithm development in electron tomography. Ultramicroscopy  \textbf{157},
  35--47 (2015)

\bibitem{veeraraghavan2010increasing}
Veeraraghavan, A., Genkin, A.V., Vitaladevuni, S., Scheffer, L., Xu, S., Hess,
  H., Fetter, R., Cantoni, M., Knott, G., Chklovskii, D.: Increasing depth
  resolution of electron microscopy of neural circuits using sparse tomographic
  reconstruction. In: 2010 IEEE Computer Society Conference on Computer Vision
  and Pattern Recognition. pp. 1767--1774. IEEE (2010)

\bibitem{wang2022dudotrans}
Wang, C., Shang, K., Zhang, H., Li, Q., Zhou, S.K.: Dudotrans: Dual-domain
  transformer for sparse-view ct reconstruction. In: Machine Learning for
  Medical Image Reconstruction: 5th International Workshop, MLMIR 2022, Held in
  Conjunction with MICCAI 2022, Singapore, September 22, 2022, Proceedings. pp.
  84--94. Springer (2022)

\bibitem{wang2021improving}
Wang, C., Zhang, H., Li, Q., Shang, K., Lyu, Y., Dong, B., Zhou, S.K.:
  Improving generalizability in limited-angle ct reconstruction with sinogram
  extrapolation. In: Medical Image Computing and Computer Assisted
  Intervention--MICCAI 2021: 24th International Conference, Strasbourg, France,
  September 27--October 1, 2021, Proceedings, Part VI 24. pp. 86--96. Springer
  (2021)

\bibitem{wang2020deep}
Wang, G., Ye, J.C., De~Man, B.: Deep learning for tomographic image
  reconstruction. Nature Machine Intelligence  \textbf{2}(12),  737--748 (2020)

\bibitem{whiteley2020directpet}
Whiteley, W., Luk, W.K., Gregor, J.: Directpet: full-size neural network pet
  reconstruction from sinogram data. Journal of Medical Imaging  \textbf{7}(3),
   032503--032503 (2020)

\bibitem{xu2020connectome}
Xu, C.S., Januszewski, M., Lu, Z., Takemura, S.y., Hayworth, K.J., Huang, G.,
  Shinomiya, K., Maitin-Shepard, J., Ackerman, D., Berg, S., et~al.: A
  connectome of the adult drosophila central brain. BioRxiv pp. 2020--01 (2020)

\bibitem{ye2018deep}
Ye, D.H., Buzzard, G.T., Ruby, M., Bouman, C.A.: Deep back projection for
  sparse-view ct reconstruction. In: 2018 IEEE Global Conference on Signal and
  Information Processing (GlobalSIP). pp.~1--5. IEEE (2018)

\bibitem{zhang2018convolutional}
Zhang, Y., Yu, H.: Convolutional neural network based metal artifact reduction
  in x-ray computed tomography. IEEE transactions on medical imaging
  \textbf{37}(6),  1370--1381 (2018)

\bibitem{zheng2022aretomo}
Zheng, S., Wolff, G., Greenan, G., Chen, Z., Faas, F.G., B{\'a}rcena, M.,
  Koster, A.J., Cheng, Y., Agard, D.A.: Aretomo: An integrated software package
  for automated marker-free, motion-corrected cryo-electron tomographic
  alignment and reconstruction. Journal of Structural Biology: X  \textbf{6},
  100068 (2022)

\bibitem{zhu2018image}
Zhu, B., Liu, J.Z., Cauley, S.F., Rosen, B.R., Rosen, M.S.: Image
  reconstruction by domain-transform manifold learning. Nature
  \textbf{555}(7697),  487--492 (2018)

\end{thebibliography}
\end{document}